\documentclass[fleqn,10pt]{wlpeerj}

\title{Feature Selection Enhancement and Feature Space Visualization for Speech-Based Emotion Recognition}

\author[1,2]{Sofia Kanwal}
\author[1]{Sohail Asghar}
\author[3]{Hazrat Ali}
\affil[1]{Department of Computer Science Islamabad, Comsats University Islamabad, Pakistan}
\affil[2]{Department of Computer Science, University of Poonch Rawalakot, Azad Kashmir, Pakistan}
\affil[3]{Collage of Science and Engineering, Hamad Bin Khalifa Univeristy, Qatar Foundation, Doha, Qatar}
\corrauthor[1]{Sofia Kanwal}{sofiakanwalgeemail@gmail.com}

\begin{abstract}
	Robust speech emotion recognition relies on the quality of the speech features. We present speech features enhancement strategy that improves speech emotion recognition. We used the INTERSPEECH 2010 challenge feature-set. We identified subsets from the features set and applied Principle Component Analysis to the subsets. Finally, the features are fused horizontally. The resulting feature set is analyzed using t-distributed neighbour embeddings (t-SNE) before the application of features for emotion recognition. The method is compared with the state-of-the-art methods used in the literature. The empirical evidence is drawn using two well-known datasets: Emotional Speech Dataset (EMO-DB) and Ryerson Audio-Visual Database of Emotional Speech and Song (RAVDESS) for two languages, German and English, respectively. Our method achieved an average recognition gain of 11.5\% for six out of seven emotions for the EMO-DB dataset, and 13.8\% for seven out of eight emotions for the RAVDESS dataset as compared to the baseline study. 
\end{abstract}

\begin{document}
	
	\flushbottom
	\maketitle
	\thispagestyle{empty}
	
	\section*{Introduction}
	
	Speech is one of the most remarkable, natural, and immediate methods for individuals to impart their emotions \cite{akccay2020speech}. The human voice reflects the emotional and psychological states of a person. The speech emotion recognition (SER) domain has been explored for more than two decades \cite{schuller2018speech}. It is  helpful in numerous applications that require human-machine interaction. SER is gaining attention in many fields like automatic speech processing, pattern recognition, artificial intelligence, and domains involving human-machine interaction \cite{lee2005toward}. By recognizing speech emotions using artificial intelligence-based methods, the interaction between humans and machines can be made more natural and effective. SER is advantageous in specific areas like automatic SER systems could help reduce accident ratio by recognizing aggressive drivers and alerting other road users \cite{ji2006probabilistic, yang2017enhanced}. In the health care field, SER systems could help physicians understand the actual emotional state \cite{harati2018depression}. Automatic learning can be improved by recognizing learners' emotions, alerting tutors, and suggesting changes to reduce negative emotions of boredom, frustration, and anxiety \cite{jithendran2020emotion}. The computer game industry may gain more natural interaction among game players and computers by exploiting audio emotion detection systems \cite{schuller2016emotion}. 
	Hence, SER has significant value in improving lives in the digital era. \\
	One of the challenging tasks in the design of a speech-based emotion recognition framework is the identification and extraction of features that proficiently describe diverse emotions and perform reliably \cite{zheng2019research}. Humans have the ability to recognize emotions using verbal and non-verbal cues. However, in the case of machines, a comprehensive set of features is required to perform emotion recognition tasks intelligently. Speech has a large range of features, however, not all of them play a significant role in distinguishing emotions \cite{ozseven2019novel} and less important features need to be discarded by dimensionality reduction methods.
	One more thing which makes defining emotion boundaries even more difficult, is the presence of wide range of languages, accents, sentences, and speaking styles of the different speakers. 
	
	 Further, there is no standardized feature set available in the literature and existing work in SER is based on the experimental studies so far \cite{akccay2020speech} and also, many of the existing SER systems are not as much accurate to be fully relied upon \cite{xu2020improve}.
	Among two broad categories of speech features: linguistic and paralinguistic, the later is more advantageous. The reason is paralinguistic features are helpful to recognize speech emotions irrespective of the language  being spoken, the text being spoken and the accent being used \cite{hook2019automatic}. Their only dependence is on characteristics of speech signals such as tone, loudness, frequency, pitch, etc. In past, many researchers used paralinguistic attributes for SER, however, a universal set based on such cues has not established yet \cite{el2011survey,anagnostopoulos2015features}. 
	Therefore, it is required to make improvements regarding the effective use of paralinguistic attributes for the SER.
	Out of various paralinguistic features, the most commonly used are prosodic, spectral, voice quality  and Teager energy operator (TEO) \cite{akccay2020speech}. Prosodic features portray the emotional value of speech \cite{tripathi2018improvement} and are related to pitch, energy and, formant frequencies \cite{lee2005toward}. Spectral features are based on the distribution of spectral energy throughout the speech range of frequency \cite{nwe2003speech}. 
	Good examples of spectral features are Mel Frequency Cepstrum Coefficients (MFCC) and linear predictor features. Qualitative features are related to perceiving emotions by the quality of voice, e.g, voice level, voice pitch and temporal structure \cite{gobl2003role}. Teager Energy Operator (TEO) based features are related to vocal chord movement while speaking under stressful situations \cite{teager1990evidence}.
	\\ 
	Some studies reported emotion recognition based on speech features combined with other modalities, for example, text \cite{lee2007emotion}, facial expressions \cite{nemati2019hybrid, bellegarda2013data} body signals, and outward appearances \cite{yu2019interactive, ozkul2012multimodal} to fabricate multi-modal frameworks for emotion analysis. However in this work, we limit ourselves to emotion recognition explicitly from speech features only.\\ 
	In this work, we use the features set of the INTERSPEECH 2010 challenge which is the combination of prosodic, spectral and energy-related features and  have a large set of attributes, i.e. 1582. Such a huge feature-space requires a good dimensionality reduction technique. One such technique most effectively used in speech research is Principle Component Analysis (PCA) \cite{jolliffe2016principal}. The efficient use of PCA requires the removal of outliers and scaling the data \cite{sarkar2014efficient}. To apply PCA appropriately, we normalized the chosen feature set and discovered 3 subsets based on the types of speech features. The PCA is applied separately to each subset. We analyzed the finalized feature set thoroughly, by drawing t-SNE graphs for the two datasets. For classification, Support Vector Machine (SVM) for One-Against-All is used. The results highlighted that appropriate selection of features and successful application of the feature selection method impacted the classification performance positively. The method achieved a decision-level correct classification rate of 71.08 \% for eight emotions using the RAVDESS dataset \cite{livingstone2018ryerson} and 82.48 \% for seven emotions using the EMO-DB dataset \cite{burkhardt2005database}. Furthermore, the proposed method outperformed the results of \cite{yang2017enhanced, venkataramanan2019emotion, tawari2010speech}. \\
	Our contributions in this work are:
	\begin{itemize}	
		\item An improved speech emotion classification method using robust features.
		
		\item An approach to apply features reduction technique on a subset of speech features.
		\item An evaluation mechanism of the selected speech features using t-SNE visualization before applying a machine learning classifier.
	\end{itemize}
	The remainder of the paper is composed as follows. Section 2 provides a review of two key modules of speech-based emotion classification framework, i.e., speech features and classifiers. Section 3 introduces the speech emotion datasets and assessment criteria utilized in this work. Evaluation results of the framework utilizing various databases and various situations are presented in Section 4. Finally, Section 5 concludes the paper. 	
	
	\section*{Literature Review}
	
	This section presents the previous work on speech-based emotion recognition with respect to selection of relevant speech features and machine learning methods used in SER systems.
	
	Yang \cite{yang2017enhanced} opted for prosodic features mainly and used mutual information as a feature selection method to choose top-80 features for the recognition task. In \cite{venkataramanan2019emotion}, few features such as MFCC, pitch, energy, and Mel-spectrogram were used, and the top-2 and top-3 accuracy of 84.13\% and 91.38\% respectively were achieved for the RAVDESS dataset. Fourier coefficients of the speech signal were extracted and applied on EMO-DB dataset in \cite{wang2015speech}, which came up with a recognition rate of 79.51\%. \cite{lugger2007relevance} used voice quality 85
	and prosodic features in combination for emotion recognition.
	In \cite{tripathi2017improvement},a speech-based emotion classification system was developed for the Bengali speech corpus. The classifiers used were ANN, KNN, SVMs, and Naive Bayes. The maximum accuracy achieved was 89\%. 
	For robust emotion classification, an enhanced sparse representation classifier is proposed \cite{zhao2014robust}. The classifier performed better in both noisy and clean data in comparison to other existing classifiers.
	In \cite{al2011novel}, a probability-based classifier, Bayesian Quadratic Discriminate was used to classify emotions. Classification accuracy of  86.67\% on EMO-DB dataset was reported by reducing the calculation cost and using a small number of features. Another study reported an ensemble classification method for speech emotion classification on Thai speech corpus \cite{prasomphan2018detecting}. The ensembled algorithms were SVM, KNN, and ANN to improve the accuracy. To make a comprehensive feature set, capable of performing well in all situations and supporting multiple languages, there is a need to include all the important speech emotion attributes. One such set of features available in the literature is INTERSPEECH 2010 paralinguistic challenge feature-set. \cite{schuller2010interspeech}. This feature-set is an extension of the INTERSPEECH 2009 paralinguistic emotion challenge feature-set \cite{schuller2009interspeech}. 
	
	\section*{Materials and Methods}
	This section discusses the material and methods used in this research.
	\subsection*{Data and preprocessing}
	The benchmark datasets used in this research are (EMO-DB) \cite{burkhardt2005database} and (RAVDESS) \cite{livingstone2018ryerson}. 
	\subsubsection*{EMO-DB}
	The EMO-DB is an acted dataset of ten experts (five male and five female) kept in the German language. It comprises seven emotion classes: anger, boredom, fear, happiness, disgust, neutral, and sadness. There are numerous expressions of a similar speaker. Ten sentences, which are phonetically unbiased, are picked for dataset development. Out of these 10 sentences, 5 sentences are short (around 1.5 seconds length) and 5 are long sentences (roughly 4 seconds term). Every emotion class has an almost equivalent number of expressions to avoid the issue of under-sampling emotion class. There is a sum of 535 expressions in this dataset. The validity of the dataset is ensured by rejecting the samples having a recognition rate less than 80\% in the subjective listening test. The metadata of the EMO-DB dataset is given in \tablename \ref{tab:PPer}.
	\begin{table}[ht]
		\centering 
		
			\begin{tabular}{l l l } 
				\hline 
				Database & \hspace{7mm}  &EMO-DB  \\ [0.5ex]
				&  \hspace{7mm} & \\ [0.5ex]
				&\hspace{7mm} & \cite{burkhardt2005database}\\ [0.5ex]
				\hline 
				No. of speakers & & 10(5 male, 5 female)\\ [1ex]
				Language & & German\\ [1ex]
				Emotions & & Anger, sadness, joy, fear \\ [0.5ex]
				& & boredom, disgust, neutral \\ [1ex]
				No. of utterances & & 535 \\ [1ex]
				Style & & Acted \\ [1ex]
				\hline 
			\end{tabular}
			\caption{\label{tab:PPer}Emotional Speech Databases (EMO-DB)}
		\end{table}
		
		\subsubsection*{RAVDESS}
		RAVDESS is multi-modular database of emotional speech and song. There are 24 professional actors each uttering 104 unique intonations with emotions: happiness, sad, angry, fear, surprise, disgust, calm, and neutral. The RAVDESS dataset is rich and diverse and does not experience gender bias, which means it has an equal number of male and female utterances, comprises a wide range of emotions, and at various levels of emotional intensity. Each actor uses two different statements with two different emotional intensities; normal and strong emotions. The total number of utterances is 1440. The information about the RAVDESS dataset is in \tablename \ref{tab:PPer1}.
		\begin{table}[ht]
			
			\centering 
			
				\begin{tabular}{l l l } 
					\hline 
					Database & \hspace{7mm}  & RAVDESS \\ [0.5ex]
					&  \hspace{7mm} &Database  \\ [0.5ex]
					&\hspace{7mm} &\cite{livingstone2018ryerson} \\ [0.5ex]
					\hline 
					No. of speakers & & 24 (12 male, 12 female)\\ [1ex]
					Language & & English\\ [1ex]
					Emotions & & Anger, sadness, calm, \\ [0.5ex]
					& & surprise, happiness, fear \\ [0.5ex]
					& & disgust, neutral \\ [1ex]
					No. of utterances & & 1440 \\ [1ex]
					Style & & Acted \\ [1ex]
					\hline 
				\end{tabular}
				\caption{\label{tab:PPer1}Emotional Speech Databases (RAVDESS)}
			\end{table}
			
			\subsection*{Preprocessing}
			The preprocessing step involved reading audio files, removing unvoiced segments, and framing the signal having 60ms length.
			
			\subsection*{Feature extraction}
			After preprocessing, we moved towards feature analysis which involved the extraction of useful features for speech emotion analysis. 
			In this research, the OpenSMILE toolbox was used and the resulting feature-set was INTERSPEECH 2010 Challenge feature-set \cite{schuller2010interspeech}. The reason for opting INTERSPEECH 2010 feature-set is because, it covers most of the features (namely, prosodic, spectral, and energy) effective for emotion recognition. The choice of initial features set is in line with the findings of  \cite{ozseven2018speech} which reported the effectiveness of these features for emotion recognition. 
			INTERSPEECH 2010 feature-set consists of a total of 1582 features. Upon a thorough analysis of the feature-set, we identified three subsets. The first subset consisted of 34 low-level descriptors (LLDs) with 34 corresponding delta co-efficients, having applied 21 functionals on each of it. This resulted in a subset of 1428 features. The second subset was of 19 functionals applied to 4 pitch based LLDs with their corresponding 4 delta coefficients resulting in a total of 152 features. The third feature-set consisted only two features, which were pitch onset and duration.
			
			\subsection*{Feature reduction technique}
			As the feature-set was huge, consisting of 1582 features, we needed a good dimensionality reduction technique. For feature reduction in SER, some of the commonly used methods are Forward Feature Selection (FFS), Backward Feature Selection (BFS) \cite{pao2005emotion}, Principle Component Analysis (PCA) and Linear Discriminant Analysis (LDA) \cite{haq2008audio}, \cite{ozseven2018content}. Among these the most commonly used and applied in many studies is PCA \cite{chen2012speech, scherer2015comparing, patel2011mapping}. PCA includes finding the eigen values and eigen vectors of the available  covariance matrix and choosing the necessary number of eigenvectors comparing to the biggest eigenvalues to create a transformed matrix. We have fed the openSMILE INTERSPEECH 2010 feature-set of 1582 features to the PCA and selected the top-100 features which we used for the classification task.
			
			\begin{figure}[ht]
				\centering
				\vspace{0.25cm}
				
				\includegraphics[width=8cm,height=6.5cm]{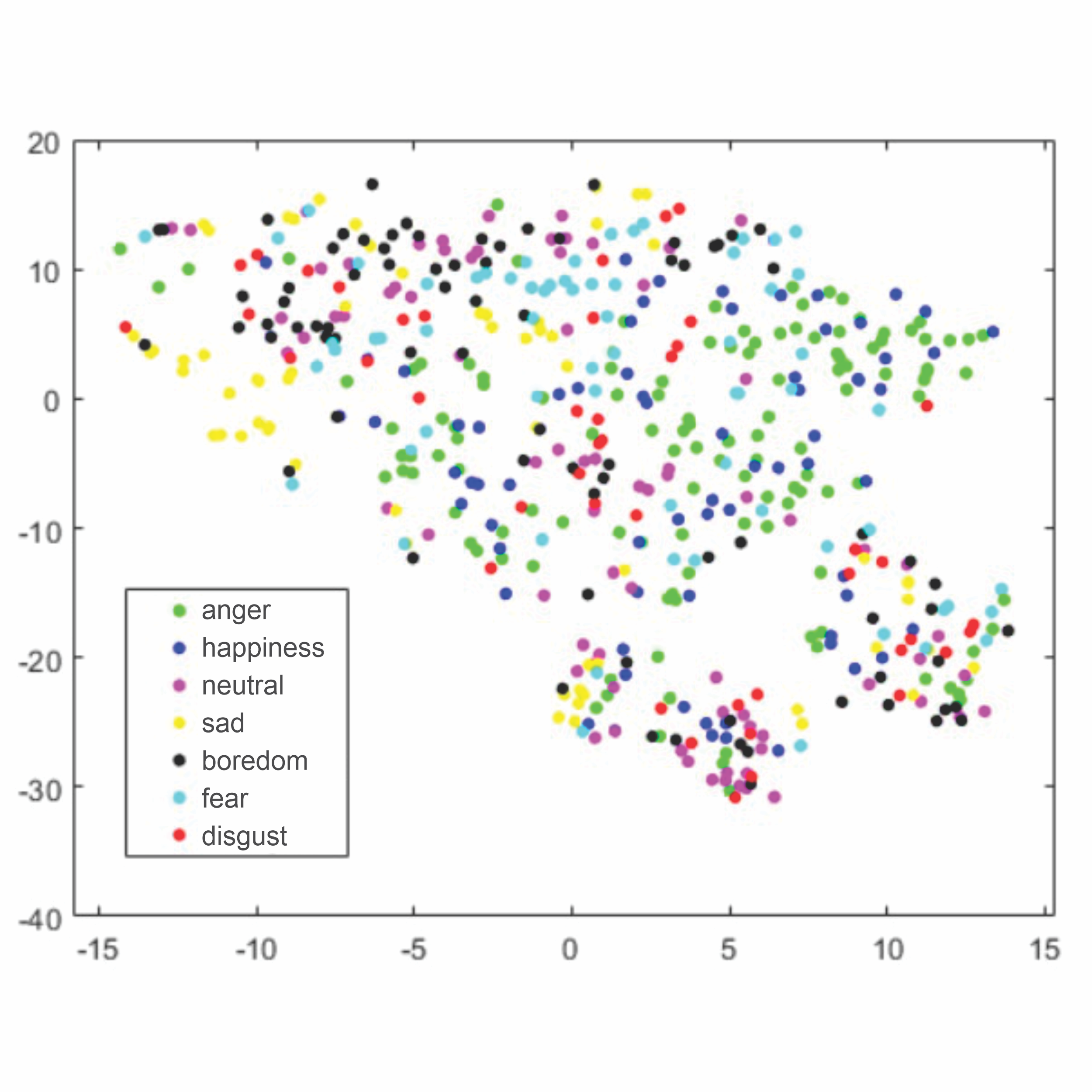}
				
				\caption{\label{fig:tsnenormalberlin} 2 Dimensional t-SNE graph of features used in \cite{yang2017enhanced} for EMO-DB dataset
				}	
			\end{figure}
			\begin{figure}[ht]
				\centering
				\vspace{0.25cm}
				
				\includegraphics[width=8cm,height=6.5cm]{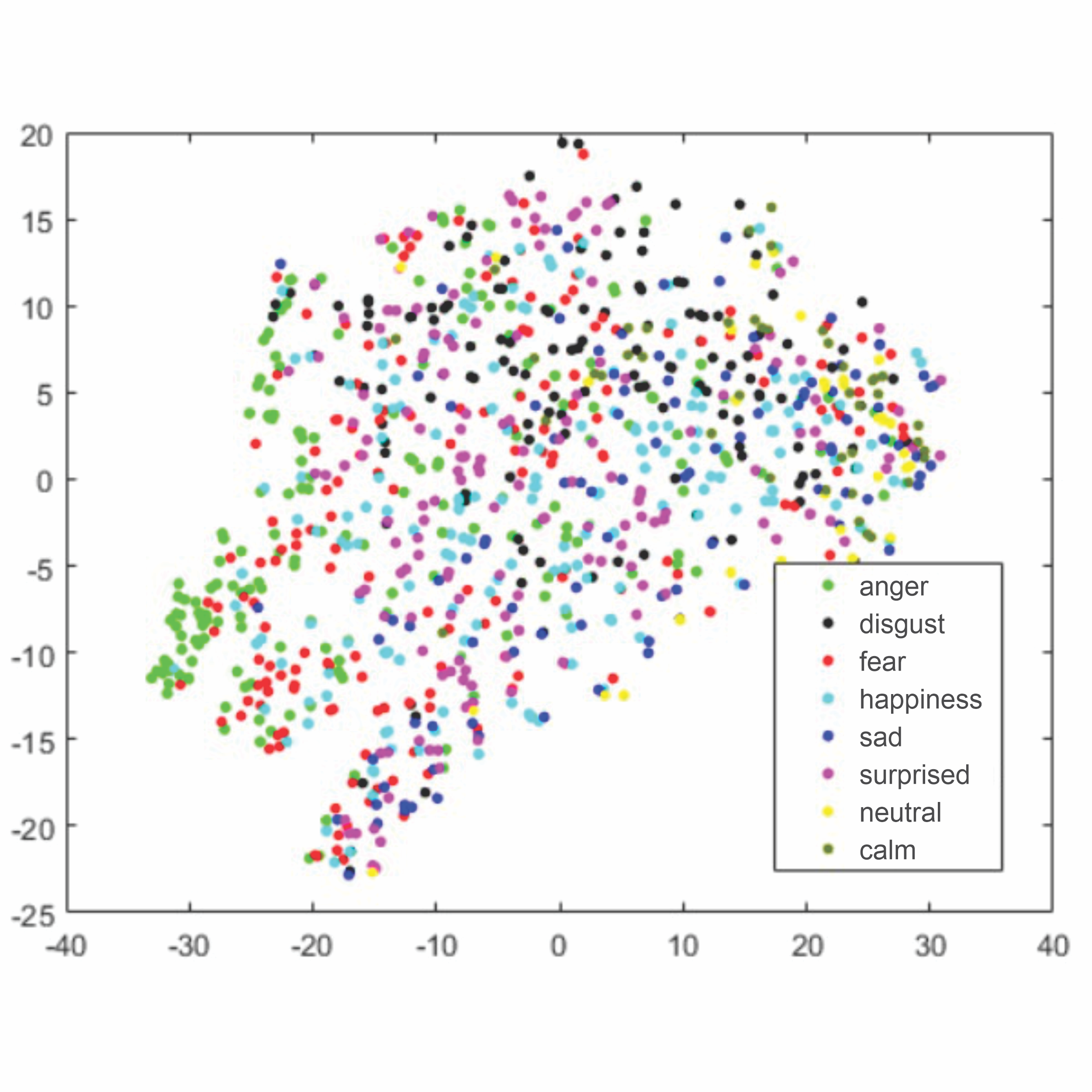}
				
				\caption{\label{fig:tsnenormalravdess} 2 Dimensional t-SNE graph of features used in \cite{yang2017enhanced} for RAVDESS dataset
				}	
			\end{figure}
			
			In \figurename{ \ref{fig:tsnenormalberlin}} and \figurename{ \ref{fig:tsnenormalravdess}}, although we can see some clusters, however, most of the emotions are evenly distributed.
			\begin{figure}[ht]
				\centering
				%
				\includegraphics[width=8cm,height=6.5cm]{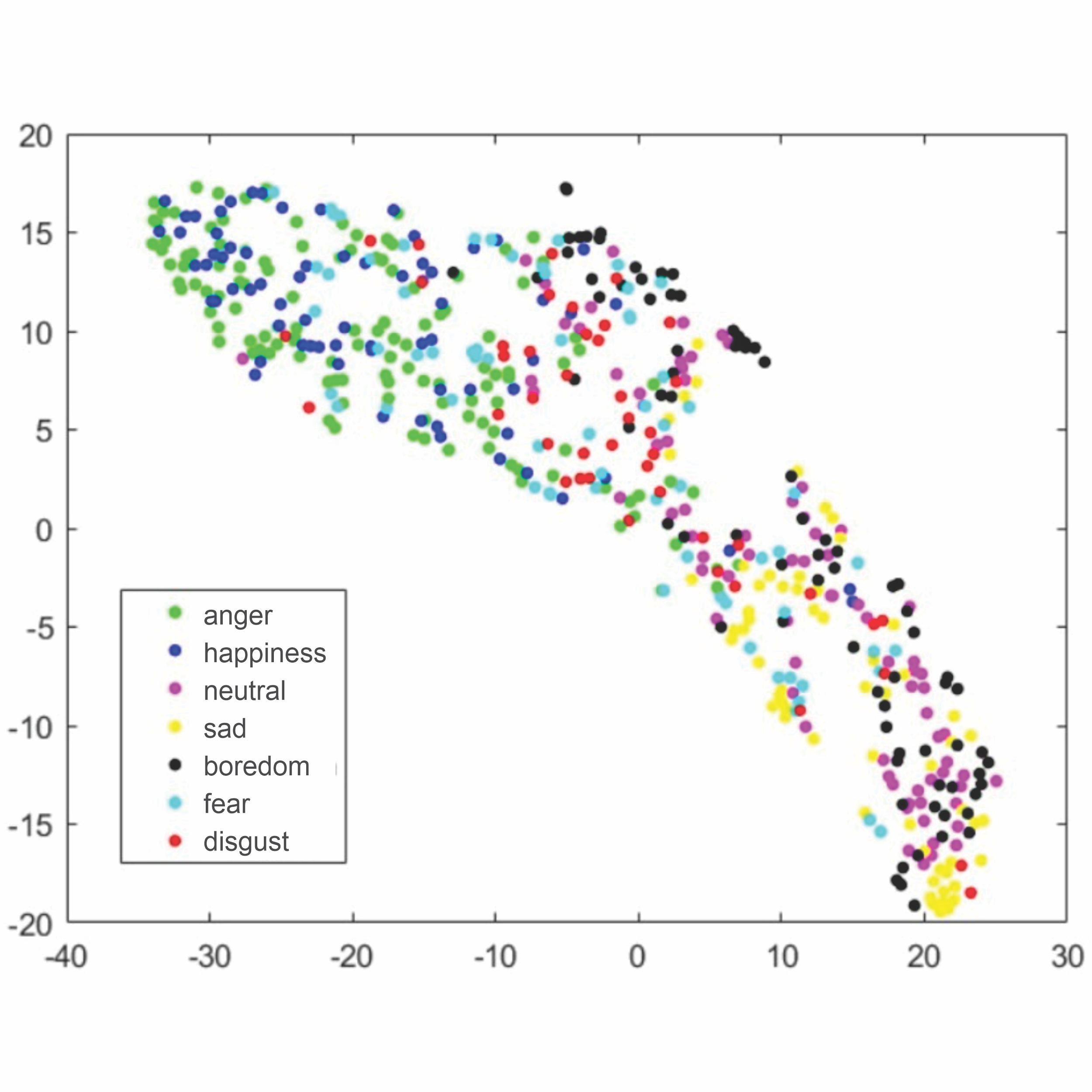}
				\caption{\label{fig:tsnenewberlin} Two-dimensional t-SNE graphs of INTERSPEECH 2010 feature-set after applying PCA for feature reduction on EMO-DB dataset.
				}	
			\end{figure}
			
\begin{figure}[ht]
				\centering
				\vspace{0.25cm}
				
				\includegraphics[width=8cm,height=6.5cm]{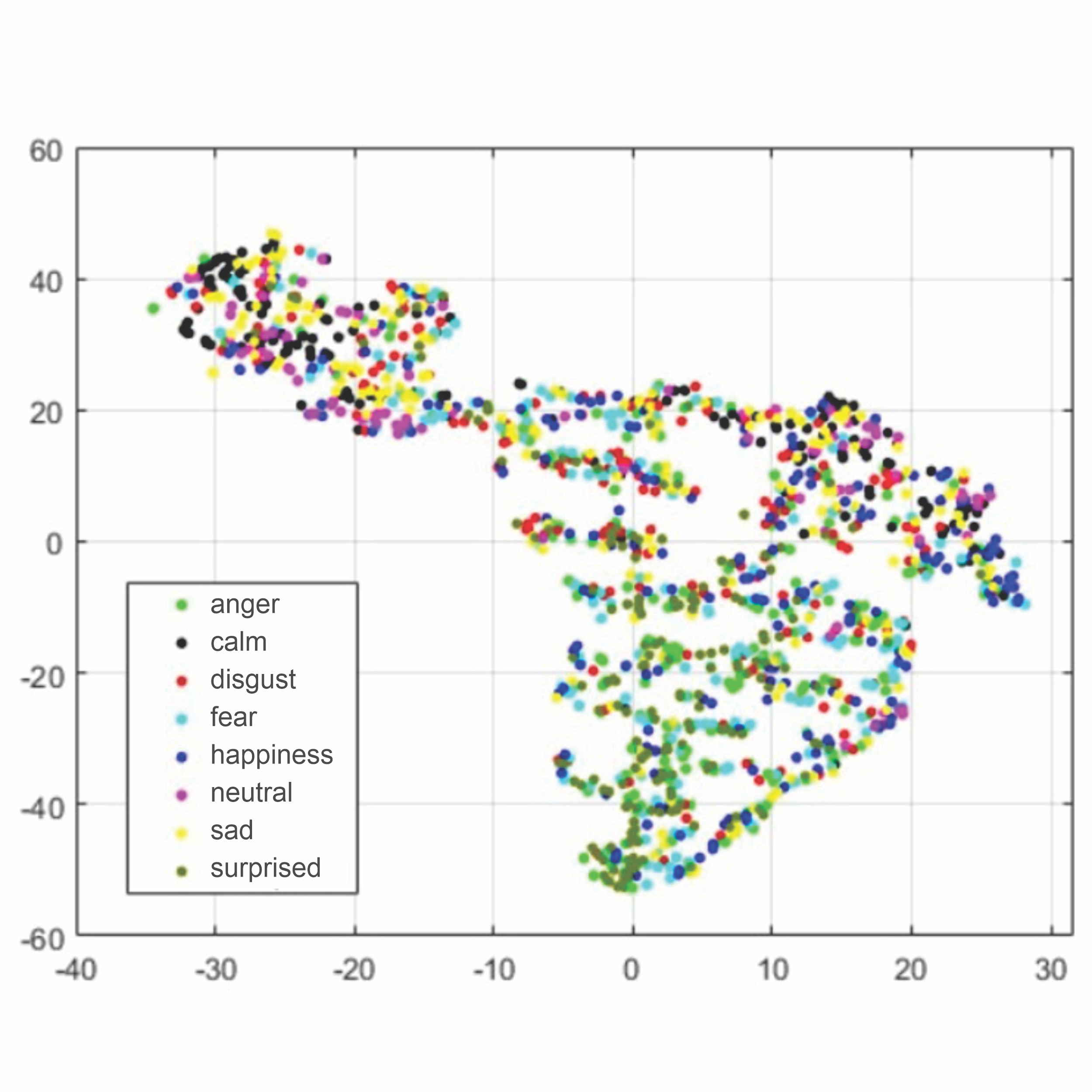}
				\caption{\label{fig:tsnenewravdess} 2 Dimensional t-SNE graphs of features after using INTERSPEECH 2010 feature-set and  applying PCA for feature reduction on RAVDESS dataset.
				}	
			\end{figure}

			As compared to feature-set utilized by \cite{yang2017enhanced}, when we used INTERSPEECH 2010 challenge feature-set and applied PCA for features reduction, a clearer representation of clusters could be seen. In \figurename{ \ref{fig:tsnenewberlin}} for the EMO-DB dataset, emotion categories of sad, neutral, and boredom were distributed in relatively compact clusters in the feature-space such that high classification accuracy was anticipated. It is also noticeable from \figurename{ \ref{fig:tsnenewberlin}}, that the data points for emotions such as happiness and anger were overlapping with each other giving a chance of misclassifying, however, overall classification of happiness and anger was not going to degrade due to being clustered at one side of feature-space. Likewise, two-dimensional representation of  t-SNE of RAVDESS dataset is shown in \figurename{ \ref{fig:tsnenewravdess}}. After using INTERSPEECH 2010 feature-set, there was a compact representation of clusters of anger, sadness, calm, and fear. Based on these observations, it was expected that overall classification accuracy for  emotion classes would greatly improve.
			
			\subsection*{Computational setup}
\begin{figure}[ht]
				\centering
				\vspace{0.25cm}
				\includegraphics[width=14cm,height=11cm]{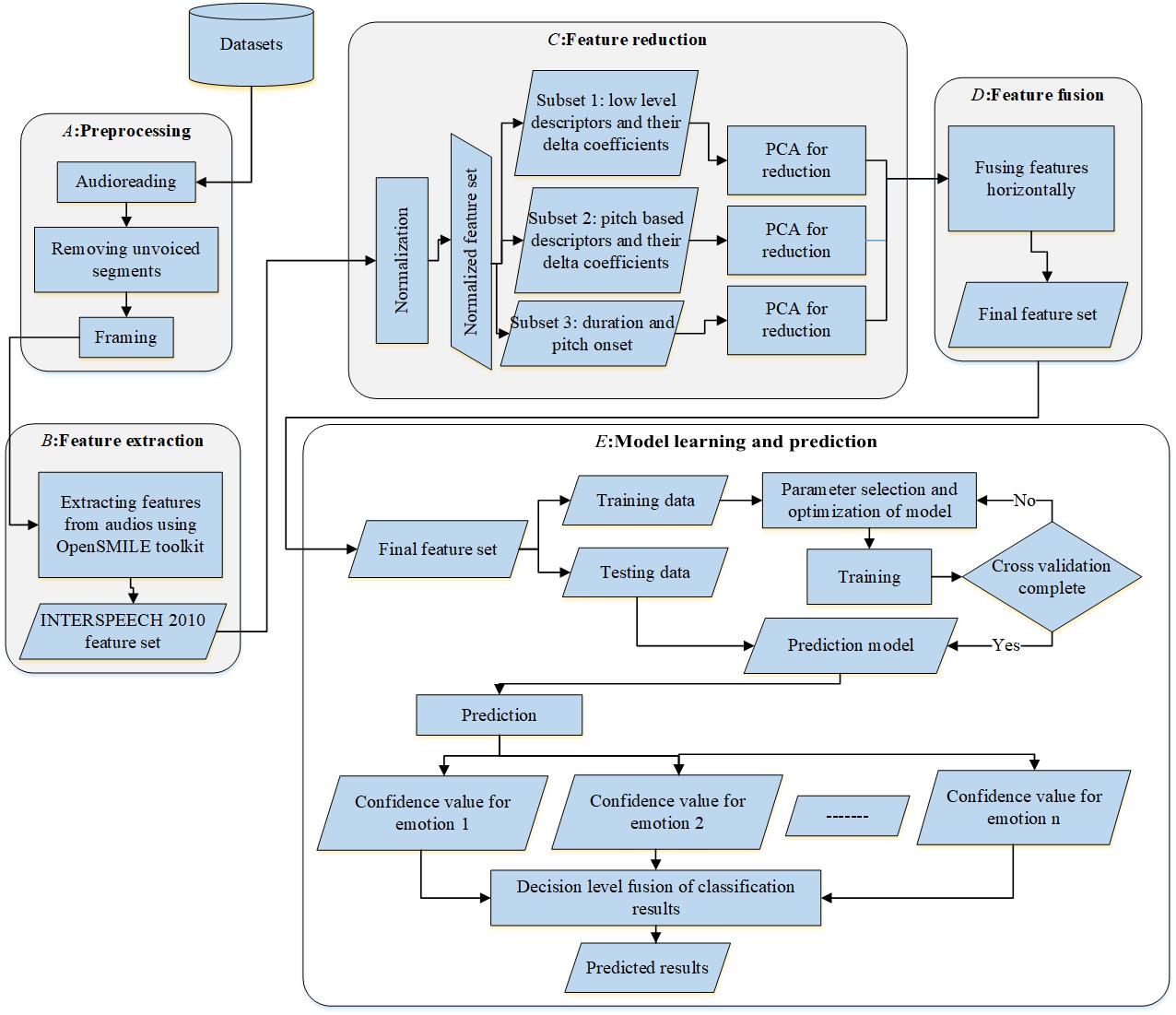}	
				\caption{\label{fig:model} The computational model is divided into modules such as A: Preprocessing, B: Feature extraction, C: Feature reduction, D: Feature fusion and, E: Model learning and prediction}
				
			\end{figure}

			To compare the performance of selected features, the process is divided into two phases: training and testing. The One-Against-All SVM classifier was trained and validated with its respective parameters to obtain the optimal model for 70\% of total samples. In the testing phase, the remaining 30\% of the samples were used to test the model. The entire framework covering preprocessing, feature selection, and model representation is visualized in \figurename{ \ref{fig:model}}.
			
			\section*{ Experimental Results and Discussion}
			
			\subsection*{Performance Evaluation Scenarios} 
			To evaluate the framework, along with metrics of accuracy and recall, two new metrics top-2 accuracy and top-3 accuracy are also introduced. Top-2 accuracy is the accuracy when the true class matches with any one of the two most probable classes predicted by the model and top-3 accuracy is the accuracy when the true class matches with any one of 3 most probable classes predicted by the model. The terminology used for evaluation is as follows: \\
			The ratio of unclassified samples over all samples in the test set is known as the rejection rate. Average classification performance for all emotions after fusion is taken as decision-level-classification-rate.
			It is defined in equation 1.\\
			\begin{equation}
				DL-Classification-rate = 
			\dfrac{DL(tp)_{m1} + DL(tp)_{m2}+......DL(tp)_{mn}}{N}
				\end{equation}
			
			where $DL(tp)_{m1}$ denotes the number of true positive utterances of emotion class $m1$ to $mn$ where $n$ is the number of emotion classes and $N$ is the total number of utterances.\\
			To evaluate the emotion classification performance of each individual emotion, the metric we are using is decision level recall rate for emotion $mi$. Mathematically it is written as given in equation 2.\\
			\begin{equation}
			DL-emotion_{i}-recall-rate = \dfrac{DL(tp)_{mi}}{DL(tp)_{mi} + DL(tn)_{mi}}
				\end{equation}
			where $DL(tp)_{mi}$ and  $DL(tn)_{mi}$ denote the decision level true positive and true negative utterances for emotion $mi$, respectively.\\
					\subsection*{Emotion Classification Performance}
			For analyzing the performance of our feature driven One-Against-All SVM based emotion classification method, the comparison with the state-of-the-art is performed. 
			After that the performance improvement of the proposed method for general test and gender dependent test is presented.
			The baseline methods considered for comparison are \cite{yang2017enhanced}, and \cite {venkataramanan2019emotion}. A summary of key features is shown in \tablename \ref {tab:comparebaseline}. In the later sections, we provide decision level emotion classification recall, top-2 accuracy and top-3 accuracy depending on which metric is provided by the reference method.
			
			\begin{table*}[!ht]
				
				\centering 
				
					\begin{tabular}{ l l l l l l } 
						\hline 
						method & Dataset & classifier & Features & Feature & Feature \\ [0.5ex]
						&  &  & &fusion & selection \\ [0.5ex]
						\hline 
						Our extended 
						
						&RAVDESS & SVM & INTERSPEECH 2010 & Yes & Yes \\[0.1ex]
						
						method &&& (prosodic, spectral&\\[0.1ex]
						
						&&& energy)&\\[0.1ex]
						
						&EMO-DB &  & & &  \\[1 ex]
						\cite{venkataramanan2019emotion} & RAVDESS & 2D CNN & pure audio, & No & No\\ [1 ex]
						& & &log mel-spectrogram & & \\ [1 ex]
						\cite{yang2017enhanced}	& LDC & SVM  & prosodic features & No & Yes  \\[1ex]
						
						\hline 
						
					\end{tabular}
				
				\caption{ \label{tab:comparebaseline} Comparison of the key characteristics of the proposed method with baseline methods of speech emotion recognition.}
			\end{table*}
			
			\subsection*{Performance comparison for general test}
			In general test performance comparison, the data samples from EMO-DB and RAVDESS datasets were used through seven-fold cross validation mechanism. In each round, one-seventh of the samples are kept for test. At the end, results are averaged over seven-folds.\\
			The work of \cite{yang2017enhanced} used LDC (Linguistic Data Consortium) dataset. The LDC Emotional Prosody Speech and Transcripts database is English database simulated by professional actors. The utterances consist of dates and numbers.  Due to unavailability of LDC dataset freely, we employed the most frequently used freely available EMO-DB dataset \cite{burkhardt2005database} and RAVDESS dataset \cite{livingstone2018ryerson} in this research. We reproduced the results of \cite{yang2017enhanced} by using both datasets and reported the comparisons.
			\subsubsection*{Comparison of average recall}
\begin{figure}[!h]
				\centering
				\vspace{0.25cm}
				
				\includegraphics[width=8.75cm,height=6.75cm]{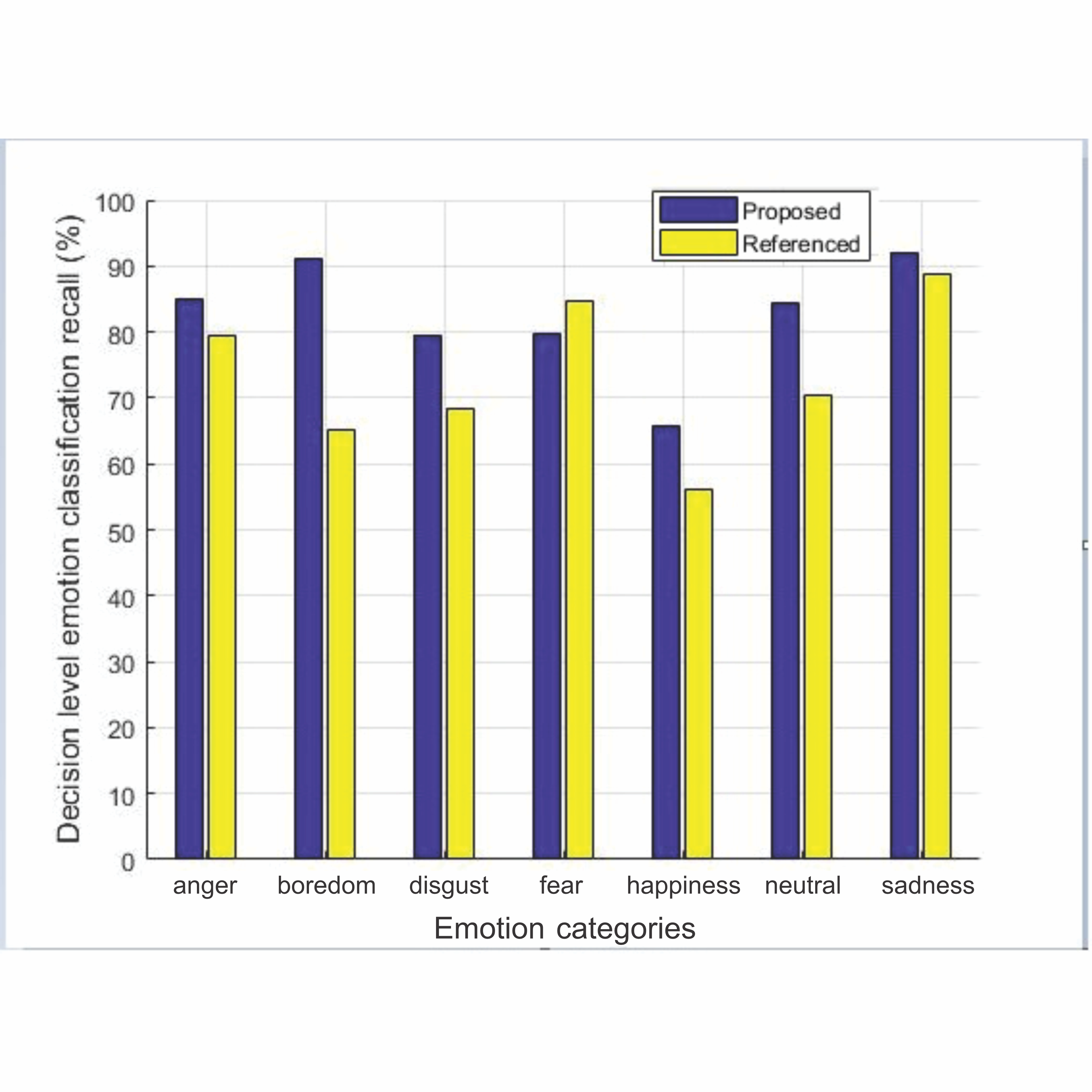}
				\caption{\label{fig:recallberlin} Decision-level emotion classification recall (\%) for each individual emotion for our method without rejecting any samples and the  method in \cite{yang2017enhanced}, using EMO-DB dataset. 
				}	
			\end{figure}

\begin{figure}[!h]
				\centering
				\vspace{0.25cm}
				
				\includegraphics[width=8.75cm,height=6.75cm]{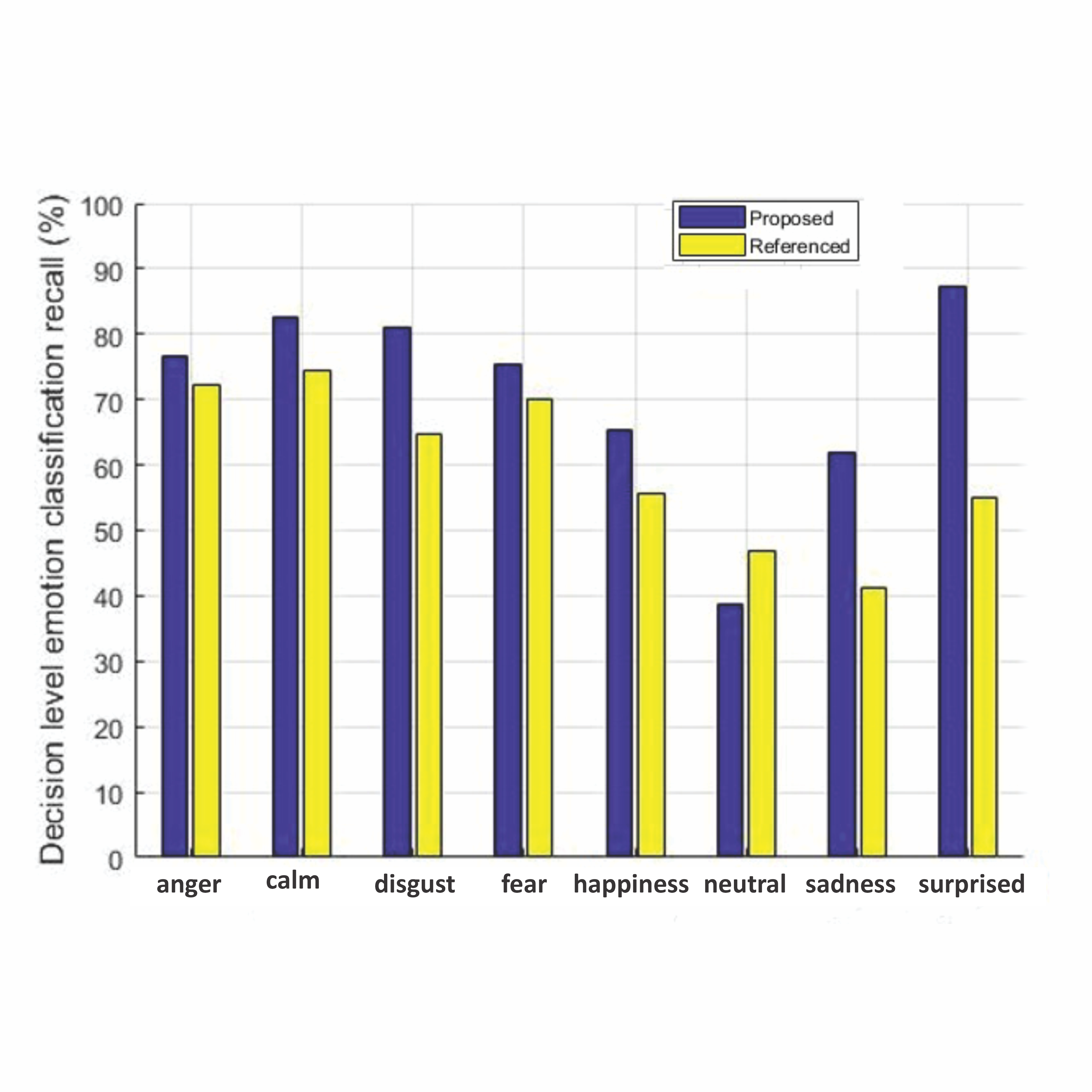}
				\caption{\label{fig:recallravdess} Decision-level emotion classification recall (\%) for each individual emotion for our method without rejecting any samples and the  method in \cite{yang2017enhanced}, using RAVDESS dataset.
				}	
			\end{figure}

			The decision level average recall for each emotion of the two datasets for our method is compared with \cite{yang2017enhanced} in \figurename{ \ref{fig:recallberlin}} and \figurename{ \ref{fig:recallravdess}}.\\
			We can see from \figurename{ \ref{fig:recallberlin}} that our method outperforms the baseline method in \cite{yang2017enhanced} for six out of seven emotions using EMO-DB dataset. In the same way performance gain for seven out of eight emotions in case of RAVDESS dataset is achieved shown in \figurename{ \ref{fig:recallravdess}}.
			The average recall values  for data samples of EMO-DB and RAVDESS datasets through seven rounds of cross validation and five upsampling are also given in Table 4.
			
			Here in \tablename \ref{tab:recalltable}, the performance gain in terms of percentage is given. Our feature selection mechanism improved the recall rate for most of the emotions on both datasets.
			\begin{table*}[!ht]
				
				\centering 
				
					\begin{tabular}{ l l l l } 
						\hline 
						Dataset & Emotion & \cite{yang2017enhanced}  & Our \\ [0.5ex]
						&  &  &  method \\ [0.5ex]
						\hline 
						
						EMO-DB

						&disgust & $68.3673
						$ & $ \textbf{79.5918
						}$  \\ [0.1ex]
						&anger & $79.4903
						$ & $\textbf{84.9206
						}$  \\[0.1ex] 
						&happiness & $56.2337
						$ & $\textbf{65.7142}
						$  \\ [0.1ex]
						&fear & $\textbf{84.6031}
						$ & $79.6825
						$  \\ [0.1ex]
						&sadness & ${88.6904
						}$ & $\textbf{92.0634}
						$  \\ [0.1ex]
						&neutral & $70.3463
						$ & $\textbf{84.4155
						}$  \\ [0.1ex]
						&boredom & $65.1515
						$ & $\textbf{91.0173
						}$  \\ [0.1ex]
						
						\hline 
						
						RAVDESS	&disgust & $64.6593$ & $ \textbf{81.1208}$  \\ 
						&anger & $72.3356$ & $\textbf{76.5117}$  \\ 
						&happiness & $55.7183$ & $\textbf{65.2828}$  \\ 
						&fear & $70.1058$ & $\textbf{75.3212}	$  \\ 
						&sadness & $41.0866$ & $\textbf{61.8233}$  \\ 
						&neutral & $\textbf{46.7948}$ & $38.6446$  \\ 
						&calm & $74.4523$ & $\textbf{82.5238}$  \\ 
						&surprised & $54.9450$ & $\textbf{87.3626}$  \\ 	
						
						\hline 

					\end{tabular}

					\caption{ \label{tab:recalltable} Performance gain of our method with \cite{yang2017enhanced} in terms of average recall for each emotion for EMO-DB and RAVDESS datasets}
				\end{table*}

				\subsection*{Performance comparison using thresholding fusion }
				Here in this section, we are doing a comparison using a thresholding fusion mechanism. Thresholding fusion mechanism is an extension of the normal classification method \cite{vapnik1998statistical}. Here it is possible to output the class label only when it has a confidence score greater than certain threshold, otherwise it is rejected. In practice, such type of classification enhancement is needed when we are only concerned with samples having the high accuracy. For example, if a psychologist wanted to know when a patient, having some psychological disorder, at certain times feels emotionally high, thresholding fusion mechanism can be used.
				\subsubsection*{General test comparison}
				
\begin{figure}[!h]
					\centering
					\vspace{0.25cm}
					
					\includegraphics[width=9cm,height=7.5cm]{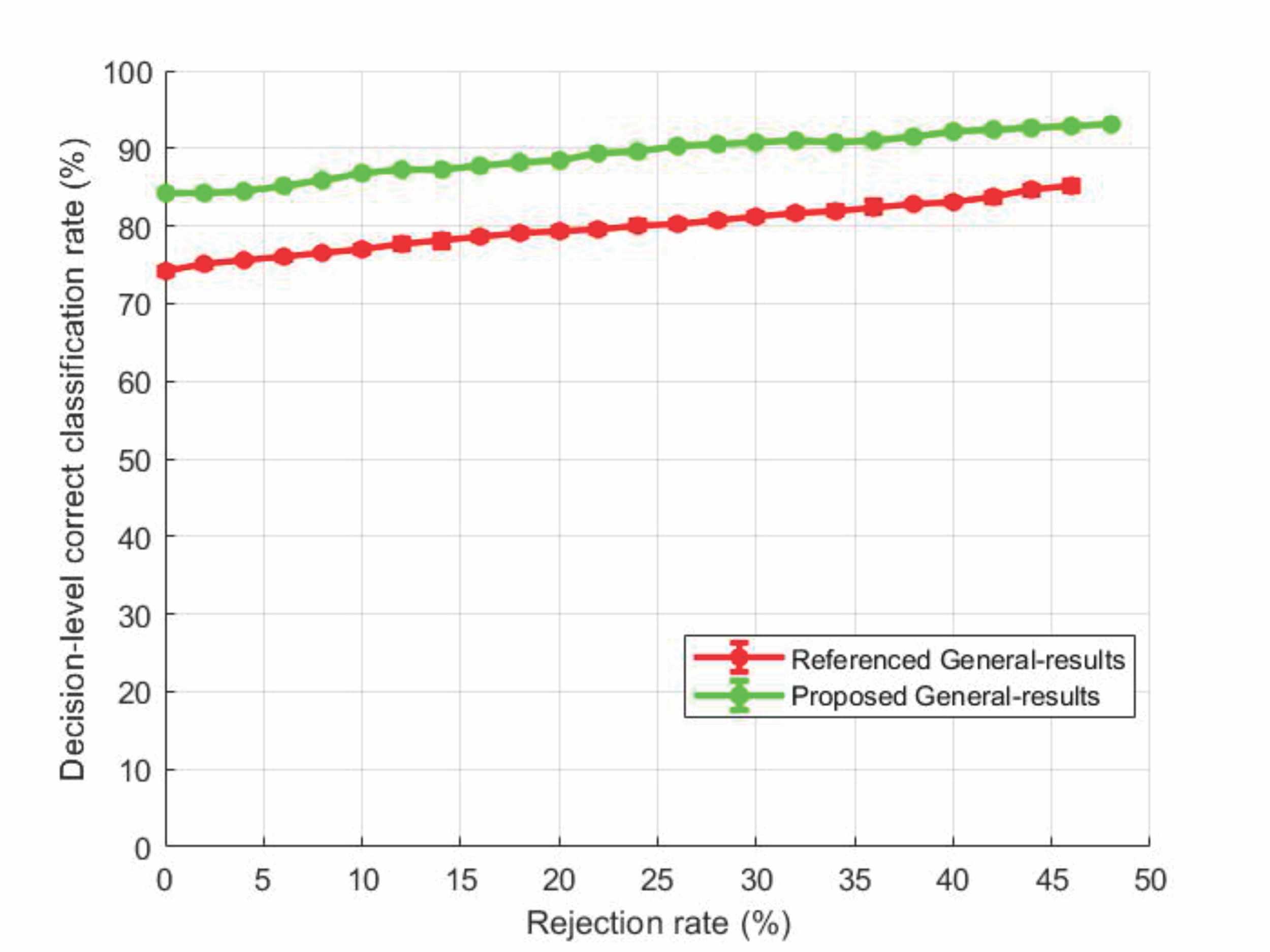}
					\caption{\label{fig:generalberlin} Decision-level correct classification rate versus rejection rate for general test of our method and the method of Yang in \cite{yang2017enhanced} using EMO-DB dataset.
					}	
				\end{figure}

\begin{figure}[!h]
					\centering
					\vspace{0.25cm}
					
					\includegraphics[width=9cm,height=7.5cm]{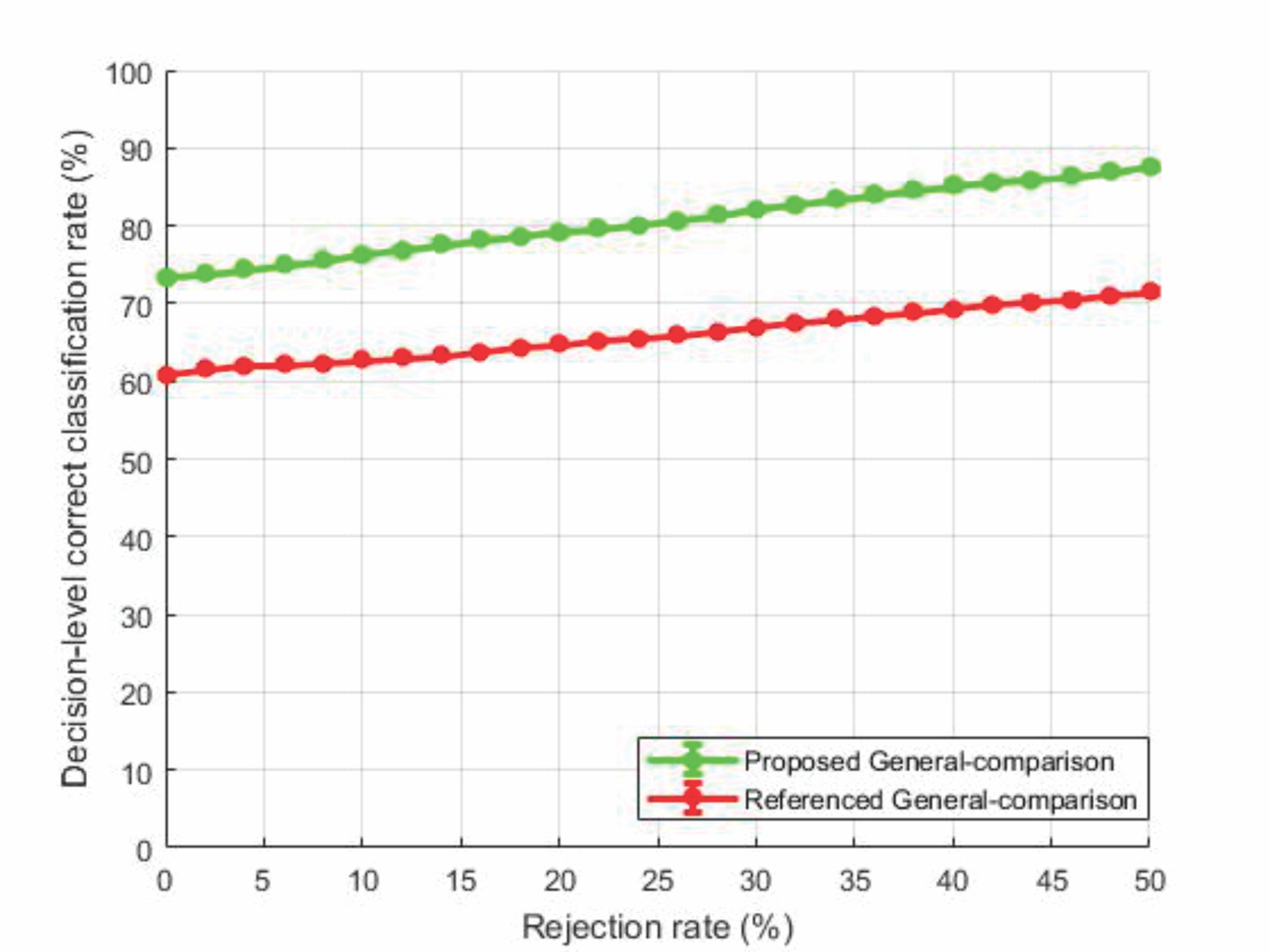}
					\caption{\label{fig:generalravdess} Decision-level correct classification rate versus rejection rate for general test of our method and the method in \cite{yang2017enhanced} using RAVDESS dataset. 
					}	
				\end{figure}

				\figurename{ \ref{fig:generalberlin}} and \figurename{ \ref{fig:generalravdess}} show the decision-level correct classification rate when samples are rejected by gauging the confidence score. 
				
				It can be seen that by rejecting about 50\% of the samples, we are able to gain above 90\% accuracy rate. The performance gain as compared to the baseline study is given in \figurename{ \ref{fig:generalberlin}} and \figurename{ \ref{fig:generalravdess}} for EMO-DB and RAVDESS datasets respectively.
				This scheme of gaining a high accuracy rate at the cost of rejecting samples could be beneficial in situations where high accuracy for precise samples is required as compared to including all samples with low accuracy.
				\subsubsection*{Gender-dependent comparison}
\begin{figure}[!h]
					\centering
					\vspace{0.25cm}
					
					\includegraphics[width=14cm,height=6.5cm]{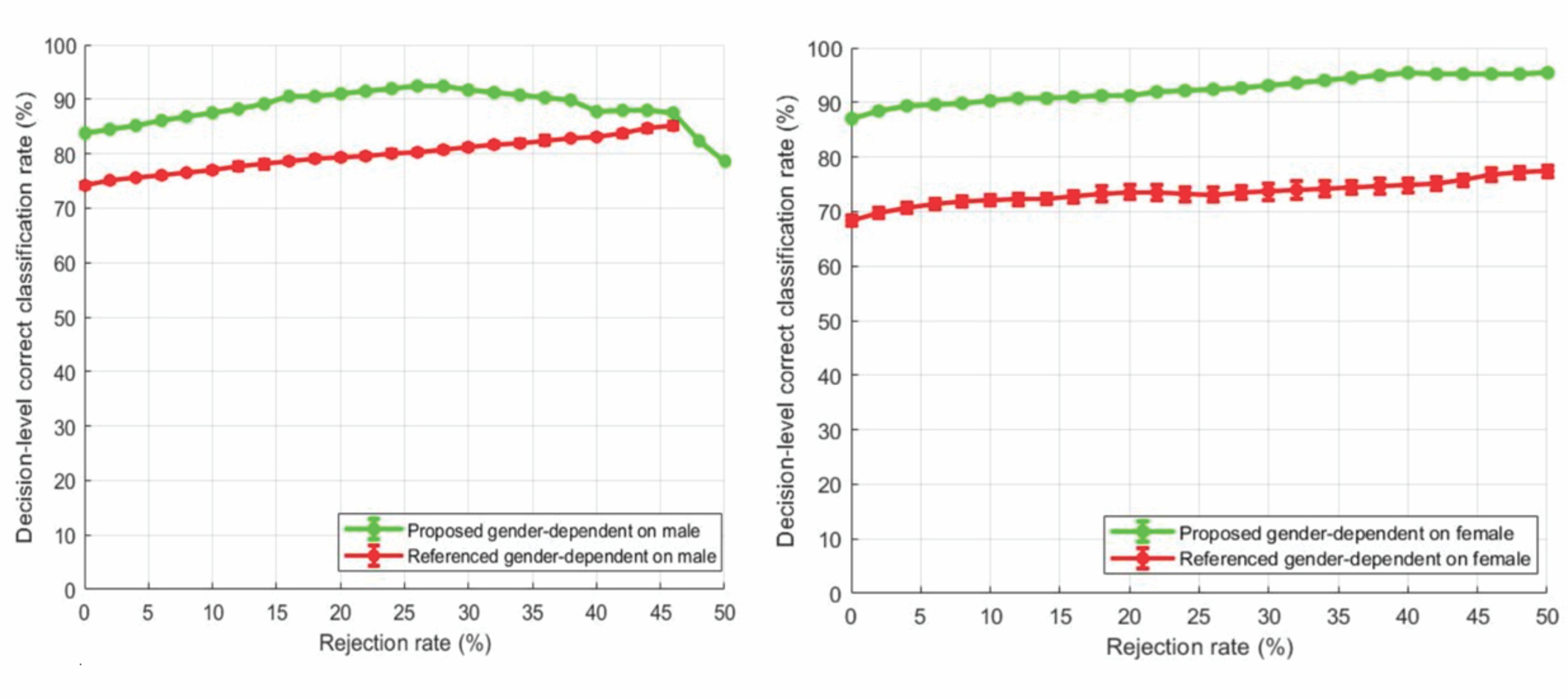}
					\caption{\label{fig:genderberlin} Comparison of decision-level correct classification rate against rejection rate for gender dependent test of our method with the baseline study \cite{yang2017enhanced}. Figure on the left is for male speakers and figure on the right is for female speakers. The EMO-DB dataset with speaker normalization, feature selection, and over-sampled training sets is used.
					}	
				\end{figure}

\begin{figure}[!h]
					\centering
					\vspace{0.25cm}
					
					\includegraphics[width=14cm,height=6.5cm]{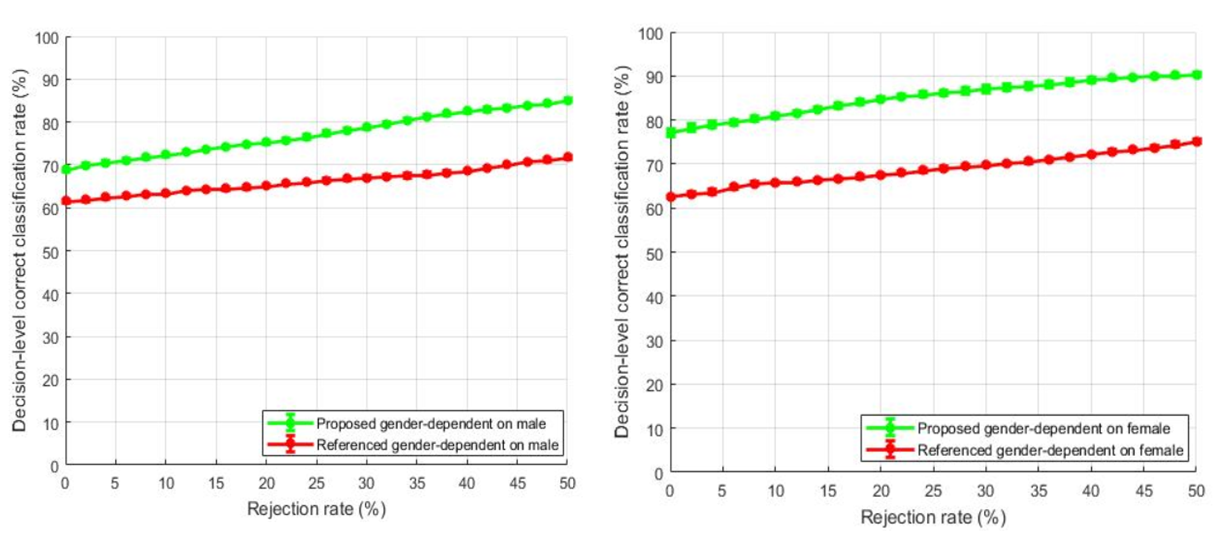}
					\caption{\label{fig:genderravdess} Comparison of decision-level correct classification rate against rejection rate for gender dependent test of our method with the baseline study \cite{yang2017enhanced} using RAVDESS dataset with speaker normalization, feature selection, and over-sampled training sets
					}	
				\end{figure}

				Decision level correct classification rate for gender-dependent tests for male and female speakers are shown in \figurename{ \ref{fig:genderberlin}} and \figurename{ \ref{fig:genderravdess}} respectively. 
				
				The features selected for gender-specific tests were based on male and female utterances and were not the same as the general tests. The performance gain of about 10\% is achieved by rejecting 50\% of the total samples.
				Our feature enhancements worked well and outperformed the baseline study of \cite{yang2017enhanced}. 
				Further, it is also observed from \figurename{ \ref{fig:genderberlin}} and \figurename{ \ref{fig:genderravdess}} that the emotion classification rate for female speakers was high as compared to male speakers. One of the reasons could be that female speakers are more accurate in decoding emotions as compared to males \cite{lausen2018gender}. The same observation is also seen in the study of \cite{yang2017enhanced}.

				\subsubsection*{State-of-the-art comparison}
			In \tablename \ref{tab:toptwoaccuracy}, top-2 and top-3 accuracy of our method for RAVDESS is compared with the one reported in \cite{venkataramanan2019emotion}. Even though, \cite{venkataramanan2019emotion} used a neural network model, the top-2 accuracy for our method improved from 84.13\% to 87.86\% having the performance gain of 3.73\%, The performance gain of 1.51\% was achieved for top-3 which is still higher then the baseline method.
			The top-2 and top-3 accuracy measures for the EMO-DB dataset are also given in \tablename\ref{tab:toptwoaccuracy}. The gender-dependent accuracy score achieved on EMO-DB dataset is also compared with the results reported in \cite{tawari2010speech}. The accuracy results averaged for both male and female speakers in our work are 85.47\%, whereas 84\% in the work by \cite{tawari2010speech}. These results indicate the effectiveness of using appropriately tuned features along with classifier.
			\begin{table*}[!ht]

				\centering 
				
					\begin{tabular}{ l l l l l } 
						\hline 
						Dataset & Metric & \cite{venkataramanan2019emotion}  & \cite {tawari2010speech} & proposed  \\ [0.5ex]
						&  &  & & extensions \\ [0.5ex]
						\hline 
						
						RAVDESS  
						&top-2 & $84.13$ & & $\textbf{87.86}$  \\ 
						&accuracy &  & &   \\ 
						&top-3 & $91.38$ & &$\textbf{92.89}$  \\ 
						&accuracy &  & &  \\ 
						
						EMO-DB
						&top-2&  & & $\textbf{92.35}$  \\ 
						&accuracy &  & &  \\ 
						&top-3 &  & &$\textbf{94.89}$  \\
						&accuracy &  & & \\
						&unweighted & & $84$ & $\textbf{85.47}$ \\
						&accuracy & & &  \\
						&(gender-wise) & &  & \\
						\hline 

					\end{tabular}
				\caption{ \label {tab:toptwoaccuracy} Performance comparison of our method with \cite{venkataramanan2019emotion} with respect to top-2 and top-3 accuracy for RAVDESS and EMO-DB datasets} 
			\end{table*}

				\section*{Conclusion}
				In this paper, we presented a strategy to improve the performance of speech emotion recognition frameworks. We used a feature selection strategy based on INTERSPEECH 2010 challenge feature-set with PCA. Here we reanalyzed and extended the existing model of multiclass SVM with thresholding fusion for speech-based emotion classification. Our in-depth analysis of feature-space through t-SNE showed that selected features represent most of the emotion-related information. The datasets used for evaluation were EMO-DB and RAVDESS having emotions of anger, calm, disgust, fear, happiness, sad, surprise, and neutral. We used PCA, where subsets of features were fed separately and finally fused to reduce the features from 1582 to 100. We used the One-Against-All SVM model for emotion classification. The performance of feature-driven SER method was compared with the baseline methods. The experimental results demonstrate the effectiveness of the selected features as the method improved the performance of speech emotion recognition. In future, it is anticipated that further investigation of various features will bring additional improvement to the recognition accuracy.

			\end{document}